# Super-performance: sampling, planning, and ecological information

**Bradly Alicea[1,2]**

## Abstract

The connection between active perception and the limits of performance provide a path to understanding naturalistic behavior. We can take a comparative cognitive modeling perspective to understand the limits of this performance and the existence of superperformance. We will discuss two categories that are hypothesized to originate in terms of coevolutionary relationships and evolutionary trade offs: *supersamplers* and *superplanners*. Supersamplers take snapshots of their sensory world at a very high sampling rate. Examples include flies (vision) and frogs (audition) with ecological specializations. Superplanners internally store information to evaluate and act upon multiple features of spatiotemporal environments. Slow lorises and turtles provide examples of superplanning capabilities. The Gibsonian Information (GI) paradigm is used to evaluate sensory sampling and planning with respect to direct perception and its role in capturing environmental information content. By contrast, superplanners utilize internal models of the environment to compensate for normal rates of sensory sampling, and this relationship often exists as a sampling/planning tradeoff. Supersamplers and superplanners can exist in adversarial relationships, or longer-term as coevolutionary relationships. Moreover, the tradeoff between sampling and planning capacity can break down, providing relativistic regimes. We can apply the principles of superperformance to human augmentation technologies.

## Introduction

Naturalistic behavior involves a tightly-integrated action-perception loop. This involves sensory sampling and an internal model that enables planning and representation, which determines the acceleration and anticipatory capacity of this loop, respectively. Enhanced acceleration and anticipatory abilities are likely the product of evolutionary specialization. Co-evolutionary relationships (in particular arms races) can result in very highly-developed abilities. This leads to a philosophical question not necessarily involving a co-evolutionary relationship: how does a fly avoid the predatory catcher (Figure 1)? In this form of the classic pursuit-evasion problem [1], flies exemplify supersamplers, while humans correspond to superplanners.

This dichotomy can also be observed in the need to generate very fast (ballistic) and very slow (finessed) movements. Examples of the former can be observed in the hand-over-hand mechanics of the slow loris [2]. Slow loris locomotion is much slower than locomotory behavior in closely-related organisms, and requires coordinated specializations in both the biomechanics and neuronal control of muscle phenotypes. These phenotypic specializations prioritize the ability to plan movements over sensory sampling, although this becomes manifest in muscle adaptations that occur alongside the ability to plan.

In organisms that produce very-fast or high-frequency movements, supersampling is necessary as a by-product of very short windows of immediate sensorimotor feedback. In these cases, observations of the supersampler requires integrating feedforward information about the environmental state, and then generating a movement fast enough to successfully match that prediction [3]. A number of trap-jaw ant species produce large amounts of mandibular force relative to their body weight (300x) [4]. This produces a movement so fast that it requires a highly accurate feedback mechanism that qualifies as superplanning. While this internal model is sufficient for prey capture, it does not transfer to jumping, as related ballistic jumping movements result in an uncontrollable jump. As computational agents, supersamplers are dominated by the rapid sampling of their sensory environment. While they do rely upon simple forms of sensory integration, the connection to movement behavior is immediate. As a result, there is little to no lag in how sensory interactions informs behavior.

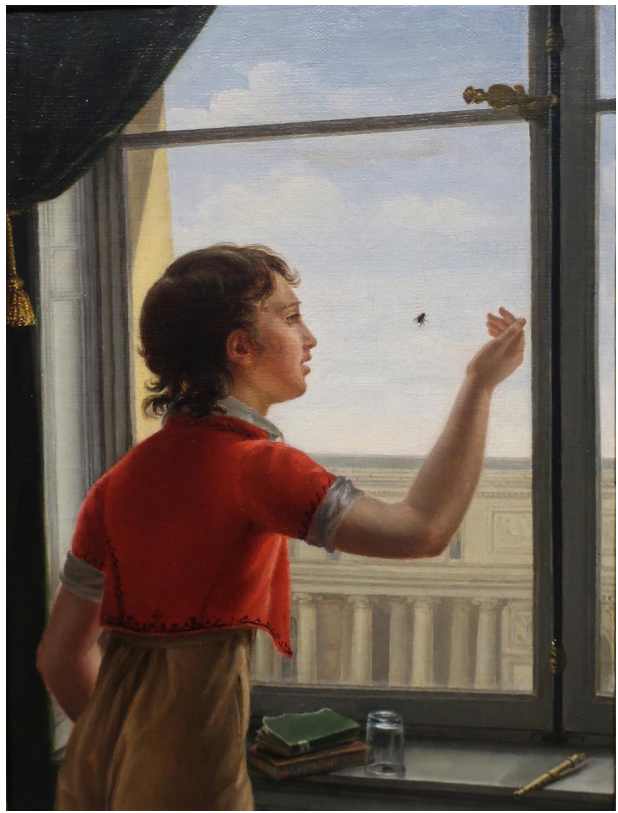

Figure 1. A visual depiction of the ability of a human to catch a fly. Painting: Isabelle Pinson, The Fly Catcher (Public Domain).

We will approach superperformance from a theoretical perspective that informs a universal cognitive model. To proceed, we define both supersampling and superplanners. At the core of superperformance is GI and the concept of ecological information processing with tradeoffs. These tradeoffs can be broken through so-called relativistic performance, which maximizes both sampling and planning. From an informational perspective, phenomena such as information aliasing and information moments provide ways to sample and interpret performance in a naturalistic context. To conclude, we consider the co-evolutionary origins of superformers modeled as a pair of complementary agents (emitter and receiver) who evaluate possibility spaces of various sampling sizes. In conclusion, we consider the application of such cognitive models to human augmentation.

**What is supersampling?**

Our first type of superperformer are *supersamplers*. Biologically, supersamplers are defined by an enhanced and hyper-specialized sensory organ. This can be either in one or multiple modalities. In the case of vision, the response sensitivity and gain properties of photoreceptors are enhanced for dark vision in nocturnal insects [5]. Another example from insect vision involves enhanced luminance sensitivity in *Drosophila* [6]. This enhanced ability allows for the organism to detect changes in luminance at shorter timescales than a visual system dominated by contrast sensitivity. Time scale plays a critical role in supersampling of the environment: there are a number of insect species with extremely high flicker-fusion frequency (FFF) rates. While human vision exhibits a 16 Hz FFF, examples from insect visual systems include FFFs of 60-100 Hz in *Drosophila hydei*, and 85-205 Hz in *Glossina morsitans* [7].

Supersampling in insects involves a number of traits working in concert to result in a set of appropriate cognitive conditions. Hyperacute vision is enabled by specialized photoreceptors that resolve target objects beyond their predicted motion-blur limit during spatial tracking [8]. When flies are in pursuit of a target, they use variables such as target size and predictions of target speed [9]. Outside of insect species, we find supersamplers of the auditory modality. Auditory supersampling requires an ability to sample and discriminate signals from a wide range of auditory frequencies. An example of this can be found in the auditory system in concave-eared torrent frogs [10]. As in the case of FFF in the visual domain, auditory samples are integrated with information from other sensory modality, which may or may not be enhanced. This will affect the sensory experience of these organisms in ways we will consider a bit later.

Returning to the trap jaw ant example [11], mandibular appendage closure is a multifunctional behavior, used for prey capture, fighting, and jumping to safety. Perhaps more interestingly, multifunctional behaviors are related to evolutionary tradeoffs and the co-option of shared and novel structures. For example, fruit flies sacrifice temporal acuity to maintain contrast sensitivity [12]. In trap jaw ants, highly-controlled movements such as drawing or balancing require a series of similar tradeoffs: slow movements with no large fluctuations in force [13].

This requires extensive co-regulation of muscle groups, and results in a highly complex sequence of physiological events [14].

**What are superplanners?**

By contrast with supersamplers, superplanners are hyper-specialized for planning based on environmental information in the form of internal models. As superperformers of the other extreme, their environmental sampling abilities can be average to poor. Superplanners will tend to exhibit high degrees of embodiment, which requires somatotopic organization [15] and multisensory integration [16]. When superplanners are implemented as computational agents, they are defined by an internal model which dominates their sensory interactions. An internal model provides a mechanism for information regulation and retention for sensory information. Biological instantiations of this internal model (somatotopic organization and multisensory integration) require a representation of the sensory world for long-term planning and movement execution. In the biological realm, superplanners will also tend towards having large brains relative to body size, which allows for the expansion of representation-rich neural structures [17]. Rather than having enhanced sensory capabilities in the temporal domain, superplanners internally store information about the environment as memories or as adaptations. Superplanning also requires enhanced spatial and/or temporal planning abilities, which in turn (and similar to supersamplers) require physiological specializations. For example, during the transition to living on land [18], such organisms required an increase in visual sensory range and related planning mechanisms as compared to ancestral species. Rather than relying solely on temporal density, superplanning involves specialized cognitive functions such as decision-making [19].

Yet other types of mechanisms might also be at play in superplanning. In the case of very slow movements, a slow down-accuracy tradeoff may exist that is inversely related to speed-accuracy tradeoffs [20]. Specializations related to muscle fibers contribute to the slowed-down movements of the turtle species *Trachemys scripta elegans* (red-eared slider). In slow tonic (SO) fibers of neck muscles critical for movement speed, SO fiber contribution to force generation is significant only in highly oxidative muscles [21]. As in the case of the slow loris, this suggests that SO fibers become predominant in a set of muscles to dampen force and power output [22], thus slowing down movement. Similar to the power amplification of muscle for very fast movement generation [23], this requires multiple regulatory mechanisms that have parallels with human technological augmentation [24].

**Interpretation**

**Ecological Information**

Gibson [25] argues that the combination of inputs, particularly covariance between input streams, results in a coherent flow of action. Gibsonian Information (GI) [26] is acquired according to a spatiotemporal Poisson distribution characterized by the parameter λ.

Environmental information is not encountered at a uniform rate: information is processed at different rates at spatially-dependent points in time, with large information moments being representative of affordances (information-rich objects) [27]. GI involves spatial information processing coupled to specific points in time, both of which are Poisson-distributed. The $\lambda$ value for a specific environment sets the gain on the information content to be processed by both supersamplers and superplanners alike. High values of $\lambda$ represents information-dense environments, with information in almost every spatially-dependent temporal moment being representative of supersampling. Low values of $\lambda$ results in long periods of sparse information. This is advantageous for the superplanner, who can utilize an internal model to fill in the gaps in direct perceptual information.

**Planning Rate/Sampling Rate Tradeoff**

A tradeoff exists between planning rate and sampling rate (see Figure 2). This planning-sampling tradeoff relates a mix of traits that enhance either sensory abilities or planning capacity in the brain. Supersamplers maintain an internal model in the form of sensory integration, and as a result their planning rate never reaches zero. In fact, for most cases this tradeoff leads to performance optima between the superplanner and supersampler regime. However, there is a relativistic regime where this tradeoff breaks down, and the planning and sampling rate are both high. In the relativistic case, sampling is both dense with respect to direct perception of the environment and the ability to draw from an internal model of this high-resolution information.

**Relativistic Performance**

This relativistic regime of performance, or in regions where the planner-sampler tradeoff is broken, is shown in Figure 2, and involves accurate predictions of the perceptual state of a supersampler by a superplanner (and vice versa). In our fly swatting example, a relativistic superplanner could put themselves in the context of a supersampler and always be successful in their attempts to catch the latter. According to this model, frogs and chameleons operate in the relativistic regime, as the muscular properties and ballistic movements of the tongue yields very large peak power outputs [28]. The introduction or elimination of sensory aliasing depends on context. Low-pass sensory aliasing results from the need to fill in perceptual gaps caused by continuous tracking of stimuli at a low FFF, while high-pass sensory aliasing can result from supersamplers sampling an informationally-sparse environment (see Figure 3a for an example).

In terms of the visual system, environmental sampling is constrained by an organism's metabolic rate [29] as well as its critical fusion frequency (CFF), or the sampling rate at which visual samples become continuous scenes [30]. Supersamplers and superplanners then deal with these constraints in different ways. In primates (who are visual superplanners), the output of MT neuronal populations are non-linear when stimuli are separated over significant periods of time [31]. In the language of GI, low λ values reinforce the need for an internal modeling mechanism to fill in perceptual sampling gaps.

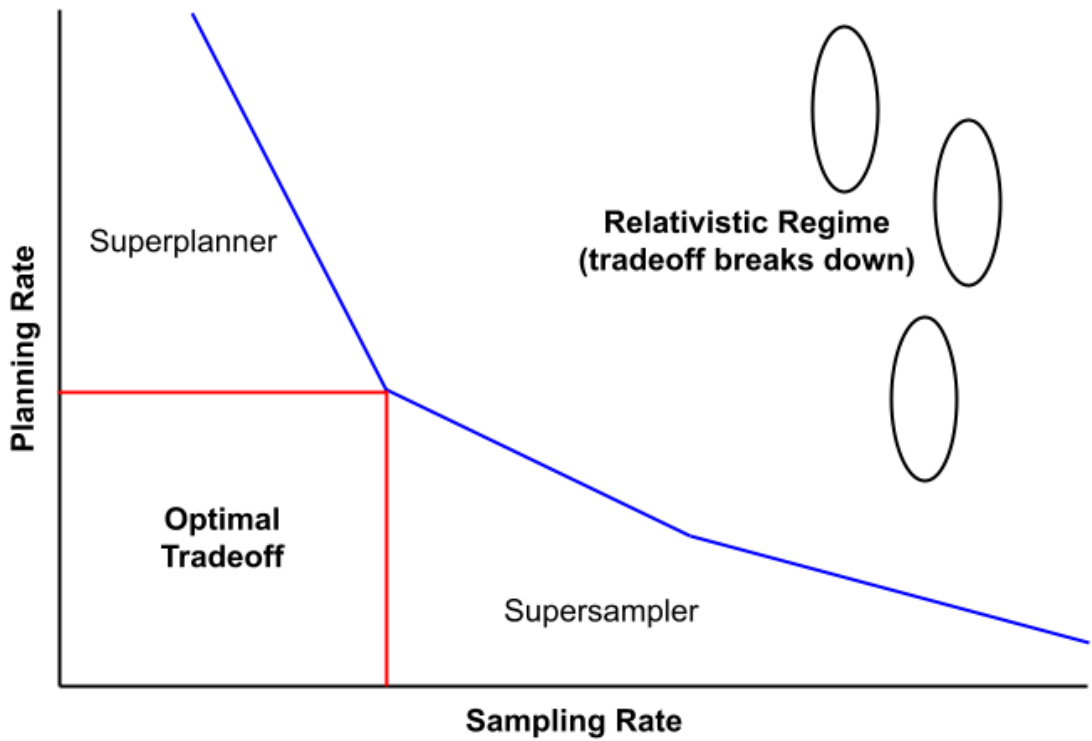

Figure 2. A theoretical graph characterizing superplanners, supersamplers, and the planning-sampling tradeoff.

**Information Aliasing**

Information aliasing (Figure 3a) occurs when normal samplers encounter situations that require the resolution of a supersampler. One example of this are flash crashes on the stock market, including the famous flash crash of 2010 [32]. Flash crashes occur when high-frequency trading algorithms trigger a massive sell-off amongst human traders in a short amount of time. The flash crash itself is not based on useful information, rather, it is triggered by spurious trends, much like deceptive environmental information. The difference in information between the superperforming agent and the normally-performing one is the source of low-frequency aliasing.

In the flash crash example, speed and quick correction are the supposed key to stability or predictability. But consider that stochastic processes such as multiscalar noise in a financial time-series can lead to transitory systems behaviors that are extreme and highly vulnerable to further cascades. In these cases, sensory aliasing without the ability to superplan favors overcorrection and the amplification of extreme or out-of-control responses. Therefore, stability can be restored through occupying the relativistic region in Figure 2.

**Active Sensing and Information Moments**

   One way to analyze superperformance is by characterizing the role of active sensation. GI relies upon active sensation in the form of continuous behaviors. In bats and weakly electric fish, an active field is emitted by the organism to detect objects within the organism's receptive field [33]. In its broader definition [34], active sensation involves active exploration of the environment with respect to environmental stimuli. The organism's sensory receptors move with its body against the environment, enabling temporal tracking and spatial exploration. It is differential movement that defines the active sensing envelope, which in turn can be evaluated using information moments (Figure 3b). Active exploration of the environment involves the modulation of sensory processing [35]. It is differential movement that defines the active sensing envelope, which in turn can be evaluated using information moments (Figure 3b). Behaviors such as reorientation involves spatiotemporal integration of intensity changes, requiring information moments to be computed by the organism [36].

   Information moments ($M_n$ in Figure 3b) are measures related to the shape of a given sensory input that represents the local density of sensory information. As supersamplers acquire high-frequency information, we should expect that the resulting information is spiky: non-uniform information should emerge in the time series are supersamplers visit high information and low information locations in space over time. These spikes can be characterized as intervals with information present at multiple scales. Another aspect of active sensation involves the origins of different behaviors used to explore their environment. In rodents, sniffing and whisking behaviors often share the same internal mechanisms [35]. In the context of superperformers, the multitude underlying mechanisms that enhance performance may involve a relatively simple set of changes.

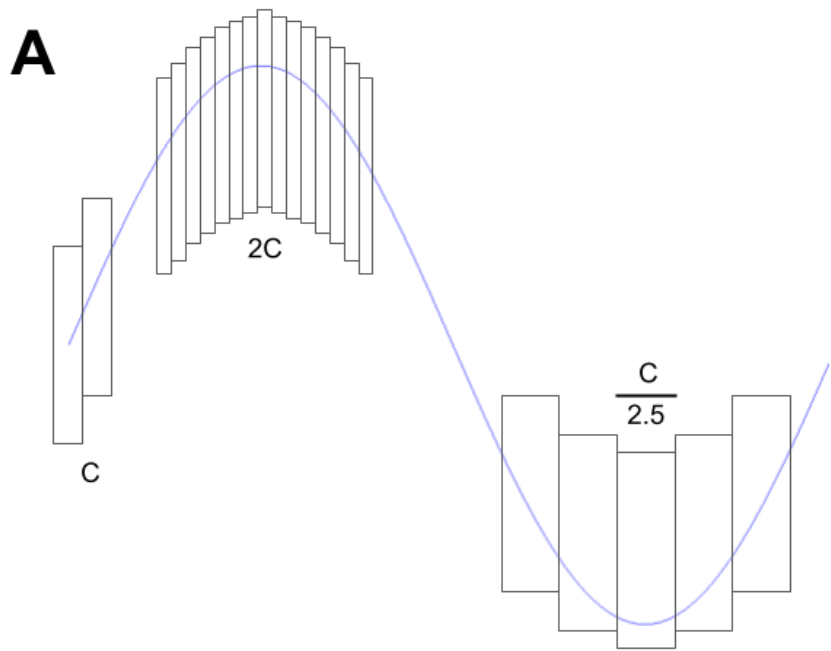

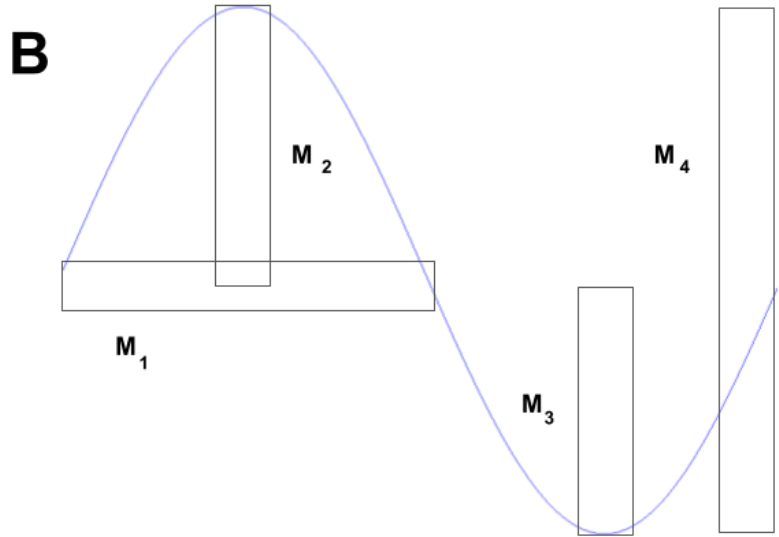

Figure 3. Examples of differences between superperformers of different types (samplers, planners) and regular performers. A: information aliasing. Sampling windows show the width of the true sampling rate (C ), width of the supersampler sampling rate (2C), and the width of the normal performance sampling rate ($\frac{C}{2.5}$). B: information moments. Moments $M_n$ demonstrate the various moments along one cycle of a sine wave.

**Co-evolutionary Superperformance**

Transferring the concept of co-evolutionary superperformance to the world of computational agents, we can model supersamplers and superplanners in terms of pairwise agent interactions (Figure 4). Our pairwise agents consist of an emitter (supersampler) and a receiver (superplanner). We begin with an environment consisting of a large ($10^n$) possibility space. Emitter agents produce patterns based on rapid sampling of the possibility space. Receivers attempt to identify the emitted patterns based on their capacity to plan possible scenarios.

Differences in the possibility space size provide us with different behaviors in the interactions between emitter and receiver. We can also consider the differences in sampling rate, along with differences in planning rate. For relatively small possibility spaces ($n > 5$), the number of patterns emitted are relatively small, and so the possibility space is exhausted rather quickly given the high sampling rate of a supersampler. Environmental sampling of any space will lead to easy detection by a superplanner (receiver). Larger possibility spaces ($n > 10$) provide a means for the superplanner to acquire more information about the emitter over a longer period of time.

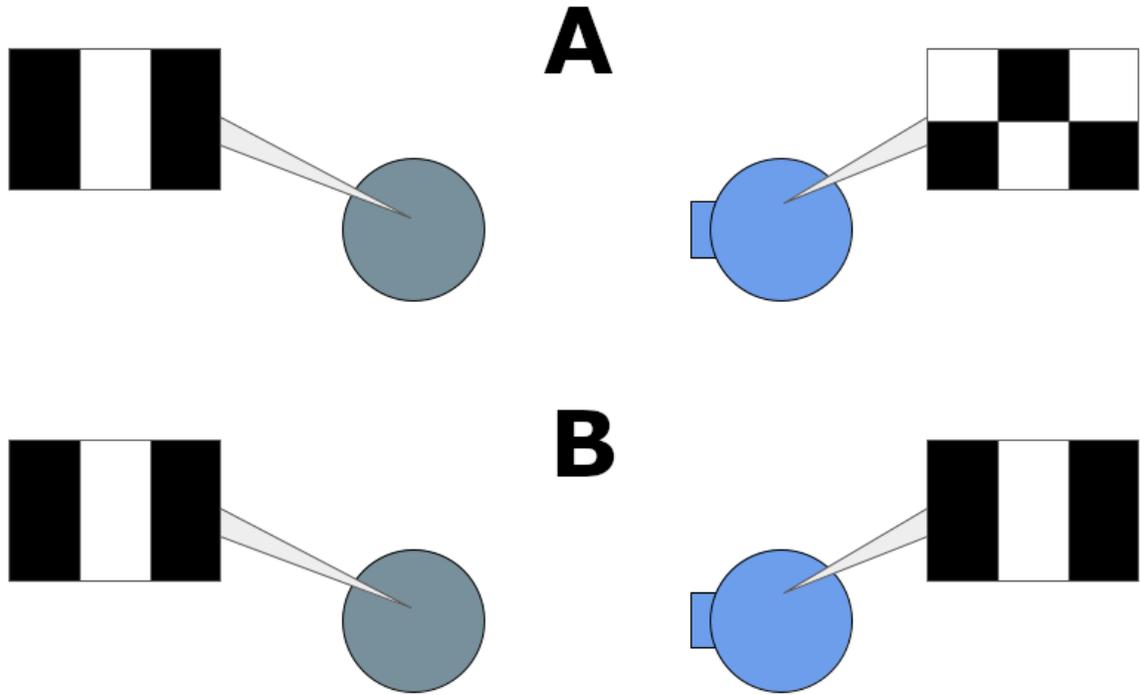

Figure 4. An example of unsuccessful and successful superperformance in a pair of agents. A: an receiver (superplanner) agent incorrectly recognizing the pattern of an emitter (supersampler) agent. B: a receiver (superplanner) agent correctly recognizing the pattern of an emitter (supersampler) agent.

## Discussion

In this paper, we consider a cognitive model of superperformance. Our focus is on two types of extreme performance: supersamplers, or cognitive systems that sample the environment at very high frequencies, and superplanners, or cognitive systems that access a large number of alternate scenarios. In nature, superplanners and supersamplers can form co-evolutionary systems that may improve the performance of both systems. Taken together, GI and tradeoffs allow us to identify phenomena such as information aliasing and information moments in our cognitive model. Another interesting feature of our cognitive model involves superperformance in the relativistic regime, where an agent can retain both supersampling and superplanning. Despite the variety of animal system examples, our cognitive model is useful for understanding and developing systems for human augmentation. Contemporary issues in Artificial Intelligence model-building are also of relevance. Superplanning resembles attempts to create world models, which attempt to define the world of a computational agent. World models simulate an internal model built through the prior sampling of data. By contrast, supersampling resembles current data-driven approaches to training deep learning models, where more data acquisition is always better.

## Superperformance Agent-based Algorithm

The lessons of superperformance can also be applied to cognitive systems design, particularly in terms of augmenting human sensory abilities. To this end, an algorithmic architecture can be proposed to enhance application of these principles to human augmentation and system design. The pseudocode for this algorithm is shown in Table 1, and is meant to be implementable in a generic computational agent. Superperformers are defined by two parameters: sampling rate $r_s$ and internal model size $I^n$. Sampling rate is the density of samples that can be operated on at time $t$. Sampling rate is constant across time, regardless of $GI$ encountered by the agent. Internal model size grows exponentially with $n$, and provides a $\tau$ substrate for memory and inference. A simulation runs with two or more agents in two classes: emitter and receiver. Each class of agent generates a new set of parameter values per agent at each fitness evaluation, thus enabling adaptive performance. Emitters emit a state at rate $r_s$, which utilizes a memoryless process that generates a phenotype given the current sensory state. The internal model of emitters is limited to a simple sensory integrator that does not process anything beyond time $t$. Receivers can store $I^n$ emitter states, along with a more elaborate integrator that integrates GI over the interval $t_n$

Table 1. Pseudocode describing the relationship between GI, agent spatial representations, and successful evasion and prediction given a simple fitness threshold.

Initialize agents ($x$ emitters, $y$ receivers).

$x(r_s)$, $y(I^n)$ provide distributions for emitter sampling rates and receiver internal model sizes, respectively.

$x_t(S)$, $y_t(S)$ is the isomorphic mapping of each agent's spatial representation.

$x_t(GI)$, $y_t(GI)$ is the Gibsonian Information available to each agent. Assume $x_t(GI) \approx y_t(GI)$.

Fitness threshold crossed when evasion or prediction criteria $< 0$. In an evasion-pursuit scenario, $r_s$ and $I^n$ are related in the following manner: increases in $I^n$ drives increases in $r_s$, whereas decreases in $I^n$ relaxes pressure on but does not decrease $r_s$.

Successful evasion is where $\int_0^t x_t(S) - y_t(S) * \tau < 0$, where $\tau$ is the lag between baseline sampling rate $y_t$ and the increased sampling rate of $x_t$.

Successful prediction is where $\int_0^t y_t(S) + I^n - x_t(S) < 0$.

Performance for both classes of agent is evaluated with a binary fitness function. For emitters, the ability to successfully escape detection from the receiver makes their behavior sustainable (fitness value of 1). Detection by the receiver (fitness value of 0) can select for higher values of $r_s$ over a series of evaluations, thus enhancing superperformance. The optimal strategy for a superperformer is not to maximize $r_s$ or $I^n$ because aliasing bias also serves to degrade performance, thus impacting the fitness function. Some forms of aliasing bias are more conducive to above-threshold fitness than others. In the case of emitters inducing aliasing in receivers, low-pass aliasing allows for emitters to escape the detection of receivers. Yet high-pass aliasing in the emitters might also negatively impact their fitness (evasion abilities) by amplifying noise instead of GI.

## Superperformance and Gibsonian Information

In the realm of superperformance, GI plays an integral role, but as understood in our definitions of supersamling and superplanning, is not solely responsible for extreme cognitive performance. For example, GI is limited in cases of temporal aliasing. We can understand the effects of temporal aliasing on GI by recalling the visual illusion of a bicycle tire rotating at frequencies greater than the human FFF. In this scenario, the tire and its spokes appear to flow backwards. This can not only result in false positives amongst supersamplers in cases of high-pass aliasing, but also leads to low-pass aliasing related misinterpretations by superplanners, as internal models misclassify ambiguities. In cases where the environment is very rich with affordance-related GI, a strategy of superplanning might out-perform supersampling, as information-dense environments require semantic discrimination that goes beyond determining structure. Indeed, the advantage provided by supersampling in insects and frogs is largely structural information (shapes and motion).

## Applications to Human Performance

We can apply the notion of superperformance to human augmentation. As demonstrated in a number of non-human species, augmented performance is the product of complex physiological traits. One can draw the parallel between superperformance in narrow ecological niches and augmentation for specific tasks in the workplace or in everyday life. As discussed in [38], characterizing physiological state with mathematical tools such as the Yerkes-Dodson curve can provide a behavioral optimum for tasks requiring optimal arousal or attention. Performance optimized around such points could offer enhancement of cognitive state, and when coupled with computer-assisted technologies (Artificial Intelligence or Head-Mounted Displays) might lead to supersampling or even relativistic performance. Utilizing methods to induce and control the direction of adaptation can also lead to augmentation leading to superperformance [24]. Wearable technologies might affect superperformance in a design-dependent manner [39], while monitoring of physiological state along with enhanced situation awareness [40] can also provide the conditions for enhanced performance and superplanning. Coupling an evolutionary algorithm to cognitive dynamics will enable future work by making comparisons with interspecies diversity in superperformance more explicit.

to cognitive dynamics will enable future work by making comparisons with interspecies diversity in superperformance more explicit.